\documentclass[nonacm,acmsmall,screen]{acmart}

\usepackage{pythonhighlight}
\usepackage{errorunderline}
\usepackage{multicol}
\usepackage{xspace}
\usepackage[capitalize,nameinlink]{cleveref}
\usepackage{subcaption}
\usepackage{wrapfig}
\usepackage{lipsum}
\usepackage{csquotes}

\Crefformat{section}{#2Sec.~#1#3}

\hypersetup{
  keeppdfinfo,
}

\begin{document}

\title{Imperative Quantum Programming with Ownership and Borrowing in Guppy}

\author{Mark Koch}
\orcid{0000-0001-8511-2703}
\email{mark.koch@quantinuum.com}
\author{Agustín Borgna}
\orcid{0000-0002-1688-1370}
\author{Craig Roy}
\orcid{0009-0002-6034-2910}
\author{Alan Lawrence}
\orcid{0009-0000-1663-7397}
\author{Kartik Singhal}
\orcid{0000-0003-1132-269X}
\author{Seyon Sivarajah}
\orcid{0000-0002-7332-5485}
\author{Ross Duncan}
\orcid{0000-0001-6758-1573}
\affiliation{%
  \institution{Quantinuum}
  \city{Cambridge}
  \country{United Kingdom}
}

\renewcommand{\lstlistingname}{Listing}

\newcommand{\TODO}[1]{{\color{red}\textbf{TODO:} #1}}

\newcommand{\Guppy}{\textsc{Guppy}\xspace}
\newcommand{\Hugr}{\textsc{Hugr}\xspace}

\makeatletter
\newcommand{\customlabel}[2]{%
  \protected@write \@auxout {}{\string \newlabel {#1}{{#2}{\thepage}{#2}{#1}{}} }%
  \hypertarget{#1}{}%
}
\makeatother

\begin{abstract}
  Linear types enforce no-cloning and no-deleting theorems in functional quantum programming. However, in imperative quantum programming, they have not gained widespread adoption. This work aims to develop a quantum type system that combines ergonomic linear typing with imperative semantics and maintains safety guarantees. All ideas presented here have been implemented in Quantinuum's Guppy programming language.
\end{abstract}

\maketitle

\section{Introduction}

Given the high cost and long queue times associated with quantum hardware, rapid debugging and iteration of quantum programs is typically not feasible.
Consequently, we rely on static techniques such as type systems to identify bugs \emph{before} programs are submitted to quantum devices.
One specific example we focus on is the application of \emph{linear types}~\cite{wadler1990linear} to statically enforce the no-cloning and no-deleting theorems of quantum mechanics.
The idea of utilizing linear typing in this context was pioneered by \citet{Selinger2006} and has since been employed as the foundation for various functional quantum programming languages.
Unfortunately, the same is not true in the setting of \emph{imperative quantum programming}, where so far, linear types have not gained much traction (with the notable exception of Silq~\cite{Bichsel2020}).
Previous attempts essentially boil down to equipping qubits with value semantics and treating quantum gates as functional primitives, effectively replicating the functional approach of \citet{Selinger2006} with more verbose syntax (see~\cref{sec:related} for a discussion).
This approach fails to capture the semantics of languages like Q\#~\cite{Svore2018,Singhal2023} or OpenQASM 3~\cite{Cross2022}, where qubits are treated as \emph{opaque pointers}, and quantum operations are implemented via \emph{side-effectful functions} acting on them.

In this work, we aim to develop a quantum type system that offers ergonomic linear typing while simultaneously capturing imperative semantics and maintaining the same safety guarantees.
To achieve this, we draw inspiration from classical systems-level programming, where similar substructural type systems have found significant success over the past decade. The most prominent example is Rust~\cite{matsakis2014rust,RustWeb}.
Unlike languages that enforce strict no-cloning rules, these languages employ linear or affine types to provide static memory safety by reasoning about pointers and aliasing.
In particular, they frame the type system in terms of \emph{ownership}~\cite{clarke1998ownership}, where certain actions on a resource (such as writing to a memory location) necessitate unique ownership of that resource.
Other parties may temporarily \emph{borrow} the resource, but must eventually return it to its original owner.
We argue that this ownership and borrowing discipline is a perfect fit for imperative quantum programming and, in fact, represents the optimal approach to implementing linear typing ideas from \citet{Selinger2006} in an imperative language.

Concretely, we develop a type system and associated ownership discipline that statically enforces the following safety properties: (1) qubits cannot be used after they are destructively measured or discarded, (2) multi-qubit gates cannot act on the same qubit, and (3) it is impossible to implicitly discard or leak qubits.
We achieve this by differentiating between operations that consume qubits (such as destructive measurement or discarding) and ones that merely act on them (such as applying quantum gates).
While a qubit cannot be used after being consumed by the former, it remains perfectly valid to reuse it after the latter.
By enforcing this distinction in the type system, we make it easy for both the user and the type checker to reason about and infer which qubits are alive at various points in the program.
Ownership, in turn, is used to constrain the places in which measurements of qubits are allowed and, on the other hand, to enforce that qubits are not implicitly discarded or leaked.
Furthermore, we demonstrate that our ownership discipline enables us to translate imperative programs (where quantum gates resemble side-effectful functions) into pure functional programs better suited for quantum optimization (see~\cref{sec:fp-translation}).

All ideas discussed in this abstract have been implemented as an update to the Guppy quantum programming language~\cite{koch2024guppy}.
In~\cref{sec:walk}, we provide an overview of Guppy and its ownership system by way of examples.
In~\cref{sec:fp-translation}, we discuss how imperative Guppy programs can be translated into a functional intermediate form.
Finally, in~\cref{sec:discussion}, we discuss related work and outline future directions.
While we explain our ideas in the context and syntax of Guppy, the ideas discussed are also applicable to other imperative quantum programming languages.

\section{Imperative Guppy by Example}
\label{sec:walk}

Guppy is a domain-specific quantum programming language embedded into Python, with a particular focus on dynamic quantum programs where classical computations (conditioned on mid-circuit measurement results) can affect the execution flow of the remaining program.\footnote{Guppy is open source and available at \url{https://github.com/CQCL/guppylang}.}
Guppy enables users to write these programs using Python's native syntax for classical operations and control flow.
However, unlike Python, Guppy is compiled ahead of time rather than being interpreted.
It features a static type system that includes Python's standard data types (e.g. \pyth|int|, \pyth|float|, \pyth|list|, etc.) along with a dedicated \pyth|qubit| type to represent qubits.
See \citet{koch2024guppy} for more on Guppy.

\subsection{Ownership \& Borrowing}

Each qubit in Guppy has an \emph{owner}, uniquely identifying the scope that can consume it by destructively measuring or discarding it.
Initially, every qubit is owned by the function scope that allocated it.
Ownership may change during program execution, but there can be only one owner at any given point in time.
For instance, consider the following function:
\begin{lstlisting}[style=mypython, numbers=left]
def bell() -> (qubit, qubit):
   q1, q2 = qubit(), qubit()  # Allocated qubits owned by this scope
   h(q1)
   cx(q1, q2)
   return q1, q2  # Transfers ownership to the caller
\end{lstlisting}
After the allocation in line 2, both qubits are owned by the \texttt{bell} function.
The call to \texttt{h} in line 3 applies a Hadamard gate to \texttt{q1}.
Notably, \texttt{h} does not acquire ownership of the qubit; it merely \emph{borrows} it temporarily.
Therefore, we know that \texttt{h} cannot destruct the qubit via measurement or discarding, so we can be sure that \texttt{q1} is still alive once \texttt{h} returns. %
Thus, \texttt{q1} can be borrowed again when we call \texttt{cx} in the next line.
We must ensure that only a single borrow of a qubit is active at any given time.
For instance, the following line would be rejected by the borrow checker:
\begin{python}
   cx(q1, $\error{q1}$)  # Error: q1 already borrowed
\end{python}
To enable this check, we also need to prevent qubits from being aliased, which we achieve via standard linearity checking on owned qubits.
Finally, owned qubits can be returned as in line 5 of \texttt{bell} function, which \emph{transfers} their ownership into the calling scope:
\begin{python}
def main():
   q1, q2 = bell()
   # Main now owns q1 and q2, so we can measure them
   assert measure(q1) == measure(q2)
\end{python}
As \texttt{main} now has ownership of the qubits, it has the right to deallocate them by measuring.
Naturally, qubits can no longer be used after they have been deallocated:
\begin{python}
   h($\error{q1}$)  # Error: q1 already consumed
\end{python}

\subsection{Borrow by Default}

Unlike in Rust, borrows in our type system are \emph{not} first-class.
It is not possible to return a borrowed value or store it in a data structure.
Consequently, borrowing more closely resembles a calling convention rather than a first-class type.
We chose this design to strike a balance between expressiveness (achieving our main goal of making imperative linear typing more ergonomic) and cognitive load (not requiring our Python user base to deal with explicit lifetime parameters).\footnote{This design aligns with Graydon Hoare's initial vision for Rust, which was later abandoned in favour of explicit lifetimes~\cite{hoare2023rust}.}

When writing a function, our default assumption is that qubit arguments are only borrowed from the caller and no ownership transfer occurs.
For example, consider the following function:
\begin{lstlisting}[style=mypython, numbers=left]
def foo(q: qubit):
   # Argument q is borrowed, not owned
   h(q)
   z(q)
\end{lstlisting}
The function argument \texttt{q} above is not owned by \texttt{foo}; it is merely borrowed from the caller.
However, \texttt{foo} is still allowed to temporarily lend this borrowed qubit to someone else -- e.g., applying \texttt{h} in line 3 -- which is called \emph{reborrowing}; analogous to the notion of reborrowing of mutable references in Rust.
This works because all reborrows expire by the time \texttt{foo} returns, ensuring that the qubit can be safely returned to the caller.
Conversely, this would not be possible if the qubit had been consumed in the interim.
As discussed before, consuming operations like destructive measurement require ownership and are therefore prohibited for borrowed values:
\begin{python}
   measure($\error{q}$)  # Error: Cannot measure qubit since it is not owned
\end{python}
If users wish to measure a passed qubit, they must explicitly assume ownership by adding an \pyth|@owned| annotation to the argument:\footnote{\pyth|@owned| syntax is inspired by Python's decorator notation and OCaml's mode syntax~\cite{janestreetOxidizingOCaml}.}
\begin{python}
def bar(q: qubit @owned) -> bool:
   # Ownership transferred from the caller
   return measure(q)
\end{python}
In turn, this means that \texttt{bar} is now also a consuming operation, and any passed qubit can no longer be used afterwards.
Also, note that the signature of the \texttt{measure} function features the same annotation: \pyth|measure(q: qubit @owned) -> bool|.

This design facilitates both the user and the type checker in reasoning about the lifetime of qubits.
The function signature explicitly indicates whether a qubit remains usable after the function returns.
In particular, by making borrowing the default, we encourage users to think about and explicitly annotate whether a subroutine deallocates an input qubit.

\subsection{Avoiding Qubit Leaks}

So far, we have mostly focussed on ensuring the safety of qubits by preventing their reuse after deallocation.
Conversely, there is also a safety concern if we forget to deallocate qubits after they are no longer in use, resulting in \emph{qubit leaks}.
Given the scarcity of qubits, this should be strictly avoided.
Programming languages like Rust address this issue by automatically discarding values from the heap once their owners go out of scope.
However, in the quantum context, we aim to avoid implicit discarding of qubits since dropped qubits often require post-selection or uncomputation to ensure algorithm correctness.
Consequently, unmeasured temporary qubits typically point to an implementation bug.
Therefore, we emit a compiler error whenever a qubit's owner goes out of scope without having measured the qubit or transferred its ownership.
In such cases, we prompt the user to make their intention explicit.
Consider the following example:
\begin{lstlisting}[style=mypython, numbers=left]
def baz(q: qubit):
   tmp = $\error{qubit()}$  # Error: Allocated qubit is not consumed
   cx(q, tmp)
\end{lstlisting}
Here, the allocated qubit \texttt{tmp} owned by \texttt{baz} would be leaked because there would be no remaining reference to it once \texttt{baz} returns.
To fix this, we could either explicitly measure \texttt{tmp} or transfer ownership of the qubit by returning \texttt{tmp} to the caller.
Additionally, note that this restriction does not apply to the qubit \texttt{q} since it is only borrowed, so the responsibility for deallocating it lies with its original owner.
In other words, while we enforce linear typing for owned qubits, we only use affine typing for borrowed ones.

\subsection{Interaction with Other Language Features}

The ownership discipline laid out so far harmoniously interacts with the other language features of Guppy.
For example, our linearity checking logic extends to arbitrary control flow (\pyth|if|, \pyth|for|, \pyth|while|, etc.) through dataflow analysis.
Furthermore, users can define custom struct types that can simultaneously contain both qubits and classical data.
Ownership and borrowing of these structs are tracked field-wise, so users can measure or transfer ownership of qubits in specific fields while the others remain valid.\footnote{This is analogous to the notion of \emph{partial moves} in Rust.}
The same principle applies to collection types like lists or statically sized arrays:
\setlength\multicolsep{\abovedisplayskip}
\begin{multicols}{2}
\begin{python}
class MyStruct:
   q: qubit
   qs: array[qubit, 42]
   x: int
\end{python}
  \columnbreak
\begin{python}
def example(s: MyStruct):
   for other in s.qs:
      cx(s.q, other)
   # Classical fields are not linear:
   s.x += 2 * s.x
\end{python}
\end{multicols}
\noindent
The \pyth|for| loop reborrows each array element in \texttt{s.qs} into a temporary variable \texttt{other} that is active within the scope of the loop body.
Note that other struct fields like \texttt{s.q} are not affected by reborrowing, so we are still allowed to borrow the qubit within the loop body when applying \texttt{cx}.
In contrast, classical struct fields like \texttt{s.x} are unaffected by linearity and can be freely copied or discarded.

\section{Functional Translation}
\label{sec:fp-translation}

One notable aspect of linear type systems, as observed by \citet{wadler1990linear}, is that they make it easier to translate pure programs into imperative ones.
Here, we utilize the converse observation: imperative programs with a sufficiently robust ownership discipline can be compiled into a pure functional equivalent~\cite{chargueraud2008functional}.
While translations of this form have been implemented in the past (the most advanced being the Aeneas~\cite{ho2022aeneas} framework capable of translating a large subset of Rust), support for control flow in these systems remains limited.
Fortunately, we benefit from the fact that our borrowing system is simpler than that in Rust and other languages, enabling us to perform a \emph{complete} functional translation.
Concretely, we compile imperative Guppy programs into HUGR~\cite{koch2024hugr}, a functional intermediate representation based on dataflow graphs designed for optimization of dynamic quantum-classical programs.
The key idea is that side-effectful functions that borrow qubits can be translated into pure functions that accept and \emph{return} these qubits.
Consider the following example:
\begin{multicols}{2}
\begin{python}
def foo_imperative(q: qubit):
   h(q)
   z(q)
\end{python}
  \columnbreak
\begin{python}
def foo_pure(q: qubit) -> qubit:
   q = h(q)
   q = z(q)
   return q
\end{python}
\end{multicols}
\noindent
The correctness of this translation is ensured by the borrow checker.
Since borrows of qubits are unique, it is safe to replace them with qubits passed by value without violating the no-cloning theorem.
Similarly, borrowed qubits cannot be measured, so it is always safe to translate the expiration of a borrow into an explicit return of the borrowed qubit.
Furthermore, since we do not permit borrowed qubits to be stored in data structures, we are certain that they will remain bound to a name within the scope of the function, making it feasible to track them during the translation.

\section{Discussion}
\label{sec:discussion}

\subsection{Related Work}
\label{sec:related}

The imperative quantum programming style is employed in various quantum languages, such as Q\#~\cite{Svore2018}, OpenQASM 3~\cite{Cross2022}, Catalyst~\cite{izaac2023catalyst}, and Silq~\cite{Bichsel2020}.
Among those, Silq is the only language that provides static safety guarantees via linear typing.
While Silq offers imperative control flow and looping, its quantum operations are functional primitives, resulting in code that closely resembles the \texttt{foo\_pure} example in \Cref{sec:fp-translation}, involving threading of linear qubits, rather than the more ergonomic \texttt{foo\_imperative} variant.
This was also the case in an earlier version of Guppy~\cite{koch2024guppy}.

Also inspired by Rust, \citet{Singhal2021} proposed a solution for statically preventing qubit cloning in $\lambda_{Q\#}$ using controlled aliasing of qubits.
However, their approach was restricted to an idealized functional version of Q\# and, adhering to Q\# semantics, did not account for ownership transfer.

\subsection{Future Work}

So far, we have only used ownership and borrowing to constrain the places in which destructive measurement of qubits is permitted.
An interesting generalisation would be to consider more fine-grained versions of borrowing that further restrict the kind of quantum operations that can be applied to qubits.
For example, a kind of borrow that only permits operations that can be described classically (i.e. ones that do not introduce superposition or phase effects) could enable automatic uncomputation in the style of Silq~\cite{Bichsel2020}.

Furthermore, to increase confidence in our type system, we want to define a core language fragment of Guppy and formally prove the safety claims made in this abstract.
In that context, the $\lambda_{\text{Rust}}$ fragment of the RustBelt project~\cite{Jung2018} and its type system could serve as an interesting starting point.
Finally, we want to prove soundness of the functional translation sketched in \Cref{sec:fp-translation}, drawing from the work done in Aeneas~\cite{ho2022aeneas}.

\bibliographystyle{ACM-Reference-Format}
\bibliography{bibliography}

\end{document}